\documentclass[pra,aps,preprint,showpacs]{revtex4}
\usepackage{amsfonts}
\usepackage{amsmath}
\usepackage{graphicx}
\usepackage{dcolumn}
\usepackage{bm}
\usepackage{epsfig}

\begin{document}
%
%

\section{Introduction}

We obtain optical instabilities in all-optical bistable systems arising from competing cooperative pathways at low input light levels. In particular for three-level atomic systems in the $\Lambda$ and V configuration interacting with two independent cavity modes, we identify the necessary conditions related to the incoherent pathways required to obtain instabilities. The instabilities arise when atomic states involved in the bistable transition are {\em leaky} and have substantial population, where the incoherent processes adversely affect the cooperative behavior of the atomic ensemble.

 Optical instabilities in atom-cavity coupled systems have been widely studied, particularly in the context of lasers and optical bistability~\cite{guido_gpa,laser-theory,haken,khitrova,arecchi_harrison}. The phenomenon of all-optical bistability (AOB) is a cornerstone effect arising from  cooperative atom-field dynamics. This phenomena has been extensively studied in order to realize all-optical switching, as well as nonlinear dynamical effects such as self-pulsing and chaos~\cite{lugiato-chaos,orozco1989}. We focus our attention to optical instabilities in AOB systems, particularly at low light levels. Traditionally the occurrence of optical instabilities arise in two well known regimes: one involves the atom coupling to multiple longitudinal modes of the cavity, and the other involves single cavity mode coupled to the atom and the instability occurs in the upper branch (also known as one-atom branch) of AOB~\cite{lugiato-chaos,orozco1989,segard}, where cooperative effect is non-existent. The former class of instabilities, arising from large number of modes, result from an interplay of a variety of time scales leading to chaos and self-pulsing. The later effect involves intense intra-cavity field that saturates the collection of atoms in absorptive AOB along with optical pumping, thus does not involve any cooperative effects. In an effort to realize optical instabilities in AOB in the regime where cooperative effects play a central role, we propose coupling two adjacent transitions in a multi-level atom to bi-chromatic fields that experience independent feedback leading to competing cooperative behavior and the resulting optical instabilities. In our previous work, we have shown the existence of instabilities and negative hysteresis in three-level ladder ($\Xi$) system with such double feedback AOB~\cite{ourpaper1,ourpaper2}. 

  Earlier, three-level atom in the  $\Lambda$-type configuration operated under Electromagnetic induced transparency (EIT) regime has been exploited to obtain instabilities using single cavity feedback~\cite{amitabh&min2005}. Recently controlled optical switching using double-cavity having a K-type multi-level system is proposed~\cite{joshi-osman}.  We seek to determine if instabilities can be obtained universally in every few level atomic configuration. It appears that the nature of the incoherent process plays a central role apart from the cooperative effects. So far the three-level atom in the V and $\Lambda$-type configurations has not been explored in the context of double feedback AOB to observe optical instabilities. However, the three-level V-system AOB has been studied in the context of spontaneous emission induced quantum interference leading to variation in the AOB switching thresholds~\cite{anton} as well as multistability~\cite{amitabh&min2003}. In absence of such quantum interference, the V system exhibits regular switching devoid of any optical instability~\cite{walls,walls2}. In this paper we study both the $\Lambda$ and V system in double-cavity AOB and are able to pin down the precise nature of the decay processes that could lead to controlled optical instabilities such as self-pulsing and chaotic output at low light levels in the cooperative branch. Such optical chaos can be used for secure communications~\cite{roy1} as well as be exploited for synchronization of chaos~\cite{roy2}. \\
    \begin{figure}[thb]
\includegraphics[width=0.45\textwidth]{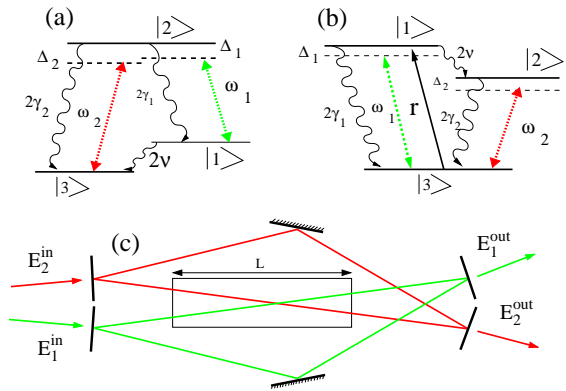}
\caption{(a) \& (b) Schematic of three-level atom in $\Lambda$ \& V configuration interacting with the two coherent fields having amplitudes $E_1$ and $E_2$ at frequencies $\omega_1$ and $\omega_2$, respectively. In case of V system, solid arrow from level $|3\rangle$ to $|1\rangle$ indicates incoherent pumping of rate $r$. (c) Two independent unidirectional ring cavities where the active medium contained within the length $L$ interacts simultaneously with the two fields.}
\label{schatomfig}
\end{figure} \\
 
  The organization of the paper is as follows, in the next section we describe the necessary theoretical model followed by results and discussion associated with the $\Lambda$ and V system, respectively. We conclude in the last section.

  \begin{figure}[thb]
\centering
\includegraphics[width=0.5\textwidth]{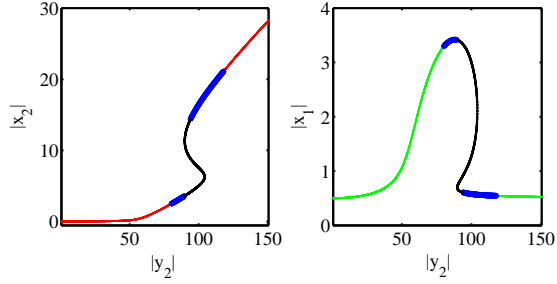}
\caption{(Color online) $\Lambda$ system: (Left \& right panel) Bistable response of output fields $x_2$ \& $x_1$ while the input field $y_2$ is varied and $y_1$ is held constant. The red and green lines indicate stable steady states, while black line indicate unstable steady states which are physically inaccessible. The thick blue line indicates unstable steady states associated with instability. The parameters are $|y_1|=0.5$,  $C_1=200$, $C_2=2000$, $\gamma_1=1$, $\gamma_2=1$, $\nu=0.1$, $\kappa_1=1$, $\kappa_2=1$, $\Delta_1=0$, $\Delta_2=0$, $\theta_1=0$, $\theta_2=5$.}
\label{lambda_bis}
\end{figure}

\section{Theoretical Model}
The schematic in Fig.~\ref{schatomfig} indicates the three-level $\Lambda$ and V configurations and the setup involving the double-cavity feedback, wherein two optical fields simultaneously interact with the atomic medium. We present the density matrix equations that govern the evolution of the atoms accompanied by the  field equations that capture feedback in the slowly varying envelope approximation~\cite{lugiato1984}. We confine ourselves to the mean-field approximation, widely used in the context of AOB, thus ignoring any spatial dependence of the field within the cavity. The propagation of the two fields inside the active medium (atomic vapor) are chosen to be co-propagating in order to eliminate the first order Doppler effect and the associated broadening~\cite{tdb}. 

 The two coherent fields having amplitude $E_1$ and $E_2$ at frequencies $\omega_1$ and $\omega_2$ couple to the atom in the $\Lambda$ configuration along the transitions $|2\rangle$ $\leftrightarrow$ $|1\rangle$ and $|2\rangle$ $\leftrightarrow$ $|3\rangle$ respectively, are shown in Fig.~\ref{schatomfig}(a). The density matrix equations are obtained under the rotating-wave approximation are given as
\begin{eqnarray}
\frac{\partial\rho_{11}}{\partial t} &=&2\gamma_1\rho_{22}-2\nu \rho_{11}+i G_1 \rho_{21}-i G_{1}^* \rho_{12} \nonumber\\
\frac{\partial\rho_{12}}{\partial t} &=&-(\gamma_1+\gamma_2+\nu-i \Delta_1)\rho_{12}-i G_{1}(\rho_{11}-\rho_{22})\nonumber\\
& &-i G_2^* \rho_{13} \nonumber\\
\frac{\partial\rho_{13}}{\partial t} &=&-\left[\nu+i(\Delta_1-\Delta_2)\right]\rho_{13}+i G_{1} \rho_{23}-i G_2 \rho_{12}~\label{finaleqn_lambda}\\
\frac{\partial\rho_{22}}{\partial t} &=&-2 (\gamma_1+\gamma_2)\rho_{22}+i G_1^* \rho_{12}-i G_1 \rho_{21}\nonumber\\
& &+i G_2 \rho_{32}-i G_2^* \rho_{23} \nonumber\\
\frac{\partial\rho_{23}}{\partial t} &=&-(\gamma_1+\gamma_2+i\Delta_2)\rho_{23}+i G_1^* \rho_{13}-i G_2(\rho_{22}-\rho_{33})\nonumber\\
\frac{\partial\rho_{33}}{\partial t} &=&2 \nu \rho_{11}+2 \gamma_2 \rho_{22} +i G^*_2\rho_{23}-i G_2 \rho_{32}\nonumber 
\end{eqnarray}
where $\Delta_1=\omega_{21}-\omega_1$ and $\Delta_2=\omega_{23}-\omega_2$ are the atomic detunings, $G_1=\vec{d}_{21}\cdot\vec{E}_1/\hbar$, $G_2=\vec{d}_{23}\cdot\vec{E}_2/\hbar$ are the Rabi frequencies, and 
$2\gamma_1$ and $2\gamma_2$ are the spontaneous emission decay rates from the level $|2\rangle$ $\rightarrow$ $|1\rangle$ and $|2\rangle$ $\rightarrow$ $|3\rangle$, respectively. The ground states decoherence arising from the non-radiative decay associated with the transition $|1\rangle$ $\leftrightarrow$ $|3\rangle$ is described by the decay rate $2\nu$.

 The three-level atoms in V configuration couple to two fields having amplitude $E_1$, $E_2$ at frequencies $\omega_1$, $\omega_2$ along the transitions $|1\rangle$ $\leftrightarrow$ $|3\rangle$ and $|2\rangle$ $\leftrightarrow$ $|3\rangle$, respectively, is shown in Fig.~\ref{schatomfig}(b). The density matrix equations in rotating wave approximation are given as
 \begin{eqnarray}
\frac{\partial\rho_{11}}{\partial t} &=&-2\gamma_1\rho_{11}-2\nu \rho_{22}+r \rho_{33}+i G_1 \rho_{31}-i G_{1}^* \rho_{13} \nonumber\\
\frac{\partial\rho_{12}}{\partial t} &=&-(\gamma_1+\gamma_2+\nu+i (\Delta_1-\Delta_2))\rho_{12}+i G_{1}\rho_{32}\nonumber\\
& &-iG_{2}^*\rho_{13}\nonumber\\
\frac{\partial\rho_{13}}{\partial t} &=&-(\nu+\gamma_1+i\Delta_2)\rho_{13}-i G_2 \rho_{12}-i G_1(\rho_{11}-\rho_{33})\nonumber\\
\frac{\partial\rho_{22}}{\partial t} &=&-2 \gamma_2 \rho_{22}+2 \nu \rho_{11}-i G_2^* \rho_{23}+i G_2 \rho_{32}~\label{finaleqn_v}\\
\frac{\partial\rho_{23}}{\partial t} &=&-(\gamma_2+i \Delta_2)\rho_{23}-i G_{2}(\rho_{22}-\rho_{33})-i G_1 \rho_{21}\nonumber\\
\frac{\partial\rho_{33}}{\partial t} &=&2 \gamma_1 \rho_{11}+2 \gamma_2 \rho_{22}-r \rho_{33} +i G^*_2\rho_{23}-i G_2 \rho_{32}\nonumber\\
& &+i G_1^*\rho_{13}-i G_1 \rho_{31}\nonumber 
\end{eqnarray}

where, $\Delta_1=\omega_{13}-\omega_1$ and $\Delta_2=\omega_{23}-\omega_2$ are the atomic detunings, $G_1=\vec{d}_{13}. \vec{E}_1/\hbar$, $G_2=\vec{d}_{23}. \vec{E}_2/\hbar$ are the Rabi frequencies, and $2\gamma_1$ and $2\gamma_2$ are the spontaneous emission rates from the levels $|1\rangle$ $\rightarrow$ $|3\rangle$ and $|2\rangle$ $\rightarrow$ $|3\rangle$, respectively. Excited states decoherence resulting from a non-radiative decay pathways from $|1\rangle$ to $|2\rangle$ is indicated with a decay rate $2 \nu$. An incoherent pump applied to transfer atomic population from level $|3\rangle$ to $|1\rangle$ plays an important role in the nonlinear dynamical behavior of V-system, whose strength is described with rate $r$~\cite{incoh}. 

\begin{figure}[thb]
\centering
\includegraphics[width=0.5\textwidth]{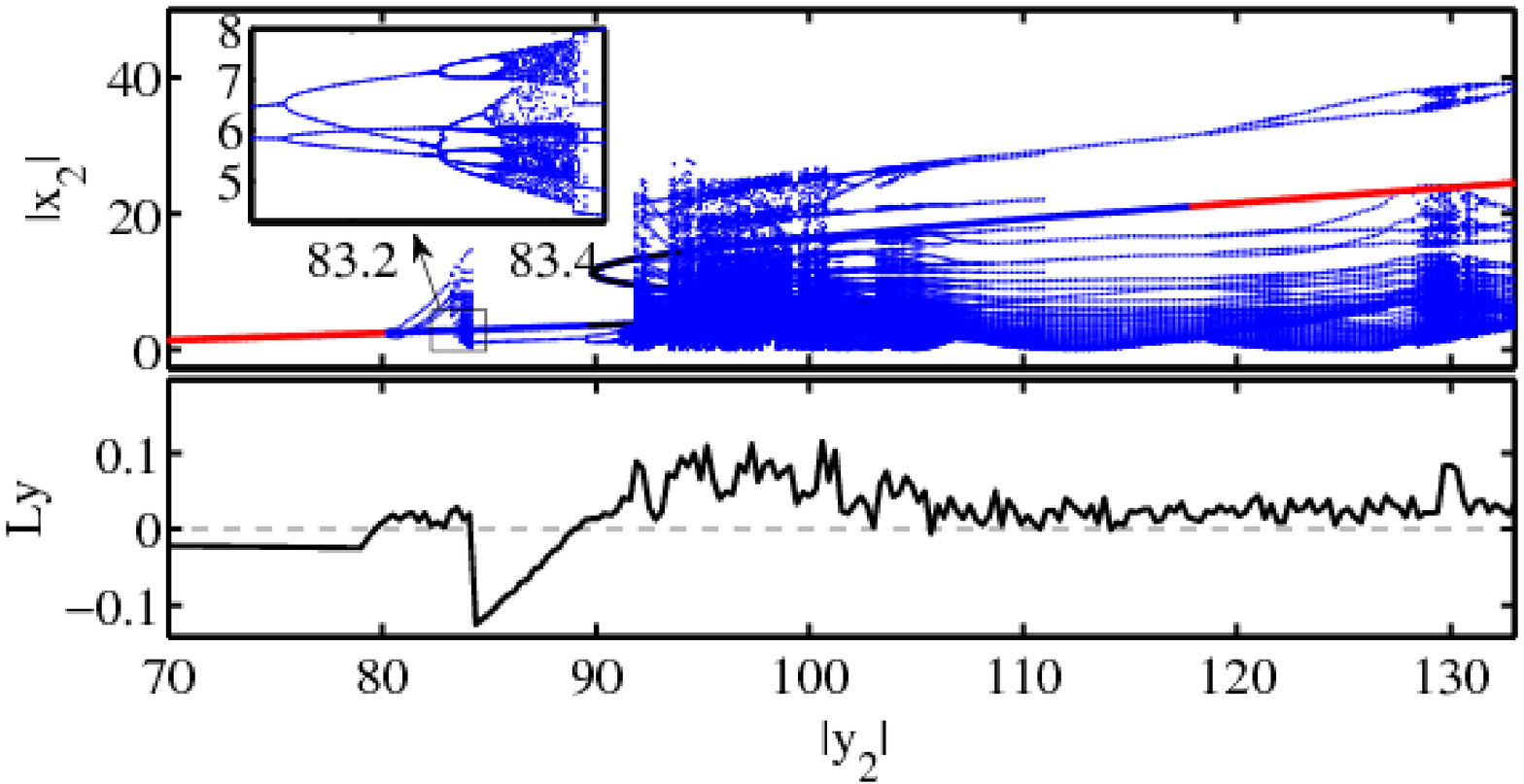}
\caption{(Color online) $\Lambda$ system: (Top panel) The bifurcation diagram corresponding to the AOB curve shown in Fig.~\ref{lambda_bis}~(left panel), where the stable steady states (red) and unstable steady states (blue) corresponds to normal switching and self-pulsed/chaotic output, respectively. The corresponding largest Lyapunov exponent is shown in the bottom panel. The parameters are same as in Fig.~\ref{lambda_bis}.}
\label{lambda_bif_ly}
\end{figure}
 
 Here, the optical fields having amplitudes $E_1$ and $E_2$ circulate within two independent ring cavities which provide sufficient feedback that leads to cooperative phenomena. The equations of motion of the fields consisting of information of boundary conditions as well as cavity feedback associated with the individual cavities are given as follows
\begin{eqnarray}
\frac{\partial x_{i}}{\partial  t}& = & \kappa_{i}\left[-x_{i}(1+i \theta_{i})+y_{i} + 2 i C_{i} \rho_{mn}\right],~\label{field}
\end{eqnarray}
where, the index `$i$' refers to the fields 1 and 2 at frequencies $\omega_1$ and $\omega_2$, respectively, and there involves normalized input fields $y_i=\vec d_{mn}.\vec E_i^{in}/\hbar\gamma_i\sqrt{T_i}$ and the output fields $x_i=\vec d_{mn}.\vec E_i^{out}/\hbar\gamma_i\sqrt{T_i}$, where $\vec E_{i}^{in(out)}$ is the amplitude of the input(output) field associated with the $i^{th}$ field that couples to the transition $|m\rangle \leftrightarrow |n\rangle$ whose dipole moment is $\vec d_{mn}$. The subscripts $m$ and $n$ denote the appropriate transition involving the atomic levels $|1\rangle$, $|2\rangle$ and $|3\rangle$, as shown in Fig.~\ref{schatomfig} (a) \& (b). The other terms in the above equation are the cavity decay $\kappa_i=c T_i/ {\cal L}_i$, the scaled cavity detuning $\theta_i=\delta_i^c/T_i$ and the cooperative parameter $C_i=\alpha_i L/2 T_i$, where $\delta_i^c=(\omega_i^c-\omega_i){\cal L}/c$ is the normalized cavity detuning, $L$ is the extent of the region which contains the active atoms, $\alpha_i$ is the absorption coefficient associated with field at frequency $\omega_i$, ${\cal L}_i$ is the total length of the corresponding cavity whose resonant frequency is $\omega_i^c$ and the transmission coefficient of the input/output mirrors is $T_i$. All the frequency units are normalized with respect to the spontaneous decay rate $\gamma_2$, unless specified otherwise. The derivation of above equations and the details of numerical techniques are along the lines discussed extensively in the Ref.~\cite{ourpaper2}.
\begin{figure}[thb]
\centering
\includegraphics[width=0.5\textwidth]{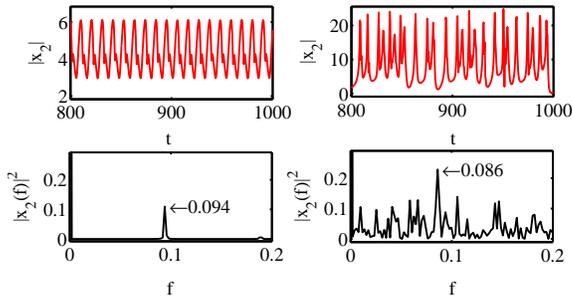}
\caption{(Color online) $\Lambda$ system: (Top) The temporal evolution indicating periodic self-pulsing and chaotic dynamics are shown in left and right panels, respectively. (Bottom) The associated frequency spectral density plots. The operating point $|y_2|=83.1$ for periodic self-pulsing and $|y_2|=95.2$ for chaotic behavior, and the other parameters are same as in Fig.~\ref{lambda_bis}.}
\label{lambda_temp}
\end{figure} 
  
 We have self consistently solved the above set of nonlinear atom and field equations in the steady state. We have undertaken a linear stability analysis, which allows us to map out the various regions of stability such as those involving stable switching, self-pulsing or chaotic output. The bifurcation diagram is also obtained in order to understand the various nonlinear dynamical features exhibited in these AOB double-cavity systems.
  \begin{figure}[thb]
\centering
\includegraphics[width=0.5\textwidth]{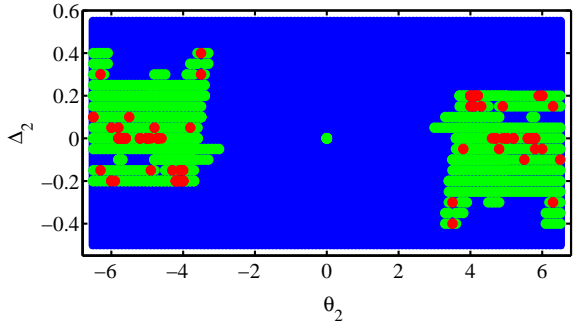}
\caption{(Color online) $\Lambda$ system: Stability domain map indicating stable-fixed-point region (blue), the self-pulsing region (green), and the chaotic region (red) as the atomic detuning ($\Delta_2$) and the cavity detuning ($\theta_2$) are varied. The other parameters are as given in Fig.~\ref{lambda_bis} with $|y_1|=0.5$ and $|y_2|=91.9$.}
\label{lambda_domain}
\end{figure}
\section{Results and Discussion}
The double-cavity AOB system interacting with three-level atoms exhibits a variety of nonlinear dynamical effects which seem to occur in all the three types of atomic configurations ($\Lambda$, V and $\Xi$ system) in appropriate regimes. It is such comparative analysis that provides insight into the optical instability and its dependence to decay channels. Earlier studies with three-level atoms have adopted a simplified approach wherein the intermediate level is adiabatically eliminated~\cite{narducci,galatola}, and the effective interaction encompasses the two-photon transition without taking into the effects of the intermediate state. This assumption effectively rules out any optical instability~\cite{Agrawal_and_Flytzanis1980,Agrawal_and_Flytzanis1981}. Such simplified treatment leads to normal stable switching. In this paper, we focus our attention on nonlinear dynamical effects associated with the cooperative (lower) branch of AOB in the $\Lambda$ and V atoms, which have not been presented earlier~\cite{ourpaper1,ourpaper2}. Our earlier study of $\Xi$ system in the double-cavity exhibits AOB, we found that the variation in the ratio of the cooperative parameter along the two adjacent transitions and the variation of the decays results in novel behavior such as self-pulsing, chaos as well as negative hysteresis.  However, the $\Lambda$ and V system are quite distinct in their behavior with regard to double-cavity AOB.

  It is important to bring out the difference between the instabilities presented here and the single mode instabilities that arise in the upper branch under strong field conditions, which are known to be well understood in the literature \cite{lugiato-chaos,orozco1989,segard}. The conventional upper branch instability invariably results from an intense nonlinear atom-field interaction and a phase mismatched feedback arising from nonzero cavity detuning. The fast nonlinear response and slow optical pumping are known to be the two competing time-scales that results in self-pulsing. In contrast, our double-cavity instabilities arise in the lower cooperative branch of AOB without involving finite cavity detuning. The cooperative branch which intrinsically involves different physical processes is responsible for the instability.

  In case of the single feedback AOB system having a multi level atomic medium, the instability associated with the cooperative branch arises from {\em leaky} population of the lower atomic level involved in the AOB transition~\cite{refr_paper}. The preparation of the atoms in the lower state such that they exhibit cooperative (or {\em collective}) phenomena is necessary to observe normal switching. A leakage of atomic population from such a coherently prepared lower atomic level leads to instability. Such an effect is easily realized in the three-level ladder system with AOB along the upper transition. The leakage of population occurs via spontaneous emission into the ground state leading to the instability in the lower cooperative branch. Such an arrangement can also occur in the $\Lambda$ system wherein incoherent decay coupling the pair of ground states is essential to obtain instability in the cooperative branch. This situation cannot arise in a V-system, as the AOB field directly couples to the ground state. Hence, V-system does not exhibit instability in the lower cooperative branch.  
\begin{figure}[thb]
\centering
\includegraphics[width=0.5\textwidth]{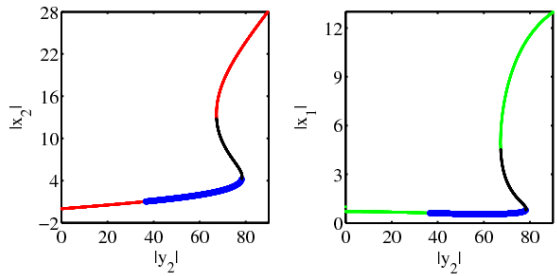}
\caption{(Color online) V system: (Left \& right panel) Bistable response of output fields $x_2$ \& $x_1$ while the input field $y_2$ is varied and $y_1$ is held constant. The red and green lines indicate stable steady states, while black line indicate unstable steady states which are physically inaccessible. The thick blue line indicates unstable steady states associated with instability. The parameters are $C_1=200, C_2=200, |y_1|=15$, $\gamma_1=1, \gamma_2=1, \nu=0.2, \kappa_1=1, \kappa_2=1,\Delta_1=0, \Delta_2=4.5, \theta_1=0, \theta_2=-2.5$, $r=2$.}
\label{v_bis_bis_2}
\end{figure}
 
 The various incoherent decay processes including spontaneous emission, relaxation of the ground state coherence as well as any incoherent pumping pathways to and from different atomic states play a critical role in dictating the nonlinear dynamical behavior of the double-cavity AOB system. Before taking the discussion further we recall some of the important results reported in our previous work~\cite{ourpaper1,ourpaper2}, which are pertinent in developing our understanding of the current system. The three-level $\Xi$ system is coupled to a pair of coherent fields that simultaneously experience feedback in two independent ring cavities, apart from normal stable switching, the cavity fields exhibit either periodic self-pulsing or chaotic nonlinear dynamical behavior in the cooperative branch. The system exhibits period doubling route to chaos. Another interesting effect is the presence of negative hysteresis at low light levels, whereas the conventional (positive) hysteresis continues to occur at higher incident intensities.
\begin{figure}[thb]
\centering
\includegraphics[width=0.5\textwidth]{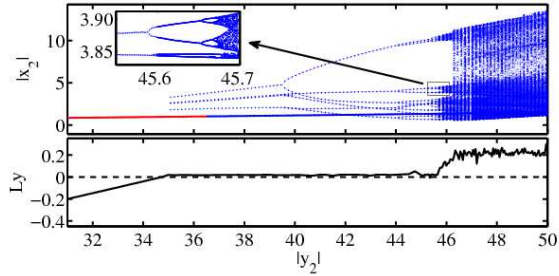}
\caption{(Color online) V system: (Top panel) The bifurcation diagram corresponding to the lower branch of AOB curve shown in Fig.~\ref{v_bis_bis_2}~(left panel), where the stable steady states (red) and unstable steady states (blue) corresponds to normal switching and self-pulsed/chaotic output, respectively. The corresponding largest Lyapunov exponent is shown in the bottom panel. The parameters are same as in Fig.~\ref{v_bis_bis_2}.}
\label{v_bif_lyap}
\end{figure}  
\subsection{$\Lambda$ - system} 
  The three-level $\Lambda$-system in a double-cavity configuration is presented here. We fix the strength of one of the input field say $y_1$ coupling the atomic transition $|2\rangle\leftrightarrow|1\rangle$ to a constant value, and then vary the other input field $y_2$ (which couples the adjacent atomic transition $|2\rangle\leftrightarrow|3\rangle$) to obtain the bistable response of $y_2$ versus $x_2$, as well as, $y_2$ versus $x_1$ as shown in the Fig.~\ref{lambda_bis}. The variation of the cavity output field $x_1$ due to the change in $y_2$ indicates the strong coupling and mutual dependence between the cavity fields $x_1$ and $x_2$ aided the atom. We have undertaken the linear stability analysis about the fixed points (steady state), and identified the unstable regions in the lower branch (indicated as thick blue line) as shown in Fig.~\ref{lambda_bis}. The usual unstable states associated with the negative slope of the S-shaped bistable response continue to exist (indicated as black line). The dynamical evolution of both the fields occur simultaneously and they mimic each other to a large extent. One can also obtain unstable states on the top branch by introducing finite cavity field detuning, akin to the instability obtained earlier~\cite{lugiato-chaos,orozco1989}. We further analyze these unstable states through the bifurcation diagram and these can be associated with either periodic self-pulsing or chaotic dynamics (see Fig.~\ref{lambda_bis} \& \ref{lambda_temp}). The stability of the periodic self-pulsing behavior is examined through the Floquet analysis whereas the existence of the chaos is confirmed through the Lyapunov exponents (Ly) which take positive values corresponding to the chaotic regime (see bottom panel of Fig.~\ref{lambda_bif_ly}). As we vary the input field $y_2$ the chaotic instability arises through a period doubling (PD) route as shown in the inset of Fig.~\ref{lambda_bif_ly}, wherein at-least one of the Floquet multipliers (Fl) crosses out of the unit circle along the negative real  axis~\cite{strogatz}, specifically for $|y_2|=83.131$ the dominant Fl is 1 and for $|y_2|=83.146$ the dominant Fl is -1.0025. 
 \begin{figure}[thb]
\centering
\includegraphics[width=0.5\textwidth]{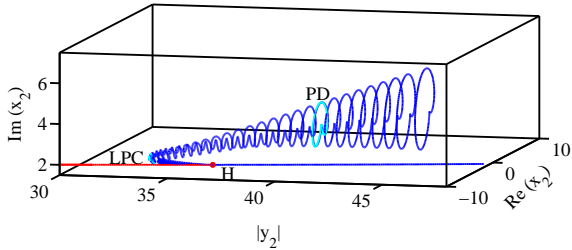}
\caption{(Color online) V system: The Limit cycle continuation from the Hopf point (H) at the very onset of instability on lower branch of AOB curve shown in Fig.~\ref{v_bis_bis_2}~(left panel). The Limit point of cycles (LPC) as well as the period doubling point (PD) are indicated as cyan colored loops. The parameters are same as in Fig.~\ref{v_bis_bis_2}.}
\label{v_limit_cont}
\end{figure}   
 
     In order to obtain instability on the top branch, it is essential to have finite cavity detuning, we have chosen $\theta_2=5$. In this regime the output amplitude of the chaotic oscillation extends from the lower branch to the top branch of the AOB response. The  coexistence of chaotic and normal stable solutions occurs for higher input field strengths. The temporal evolution and the frequency spectrum of the cavity output fields for two input field strengths, as shown in Fig.~\ref{lambda_temp} for two different $|y_2|$ values, giving rise to periodic self-pulsing and chaotic behavior. The presence of the double feedback plays a crucial role in creating these instabilities. This instability is suppressed if the feedback for one of the fields is removed. The competition between atom-field nonlinear interaction along the two cooperative branches and the associated decay pathways within the atom play a critical role in obtaining these instabilities which are quite robust and occur over a wide parameter range. The parameter space associated with this system is extremely large with fourteen independent physical parameters involving decays, detunings and couplings. In order to steer the system systematically across these variety of instabilities we take recourse to stability domain maps, between two system parameters such as detunings $\theta_2$ and $\Delta_2$ and are shown in the Fig.~\ref{lambda_domain}. One can identify islands of chaos (red) within the periodic self-pulsing domains (green) as well as, regimes of normal switching (blue). 
       
\subsection{V - system}
  We would like to explicitly bring out the importance of the incoherent (decay) pathways within the atom that are crucial to the occurrence of instability. The absorptive AOB relies on {\em collective absorption} in the {\em cooperative} (lower) branch, in contrast to a saturation of independent atoms in the one-atom (upper) AOB branch. Thus, the preparation of the {\em collective} state in the atomic ground state is crucial for realizing regular switching in AOB. The incoherent processes that directly affect this {\em collective} preparation of the atoms leads to instability such as self-pulsing and chaos. A slow time scale can be associated with the decay of the initial preparation of the atom in ground state whereas nonlinear interaction  of the atoms and the fields within the cavity provides the fast time scale, and a competition between these timescales leads to the self-pulsing behavior. We note that in the $\Xi$ and $\Lambda$ configurations the intermediate state decay and the ground state decoherence, respectively, provide the incoherent decay pathways that turns the state responsible for collective absorption {\em leaky}, leading to instability. However, in the V system the shared ground state across the two transitions can not be {\em leaky} and does not lead to any instability in the lower (cooperative) branch of AOB. Essentially, it behaves like a pair of two-level atoms interacting independently with coherent cavity fields, coupled to a common ground state. Thus, in order to produce instability we again address the {\em collective} state and introduce an incoherent pump between the ground state to one of the excited states ({\em i.e} from $|3\rangle$ to $|1\rangle$ as shown in Fig.~\ref{schatomfig}(b)). Introduction of this incoherent pathways leads to unstable steady states  over the lower branch of AOB curve, and is shown in Fig.~\ref{v_bis_bis_2} (indicated with thick blue line) and is essential for obtaining optical instability. The associated  bifurcation diagram is given in the Fig.~\ref{v_bif_lyap}, and again indicates the transition of the cavity output field from normal stable switching to chaos through a period doubling route, as shown in the inset. We also determined the Lyapunov exponents (see Fig.~\ref{v_bif_lyap} bottom panel) to demarcate the range of input field $|y_2|$ which lead to these instabilities. 
\begin{figure}[thb]
\centering
\includegraphics[width=0.5\textwidth]{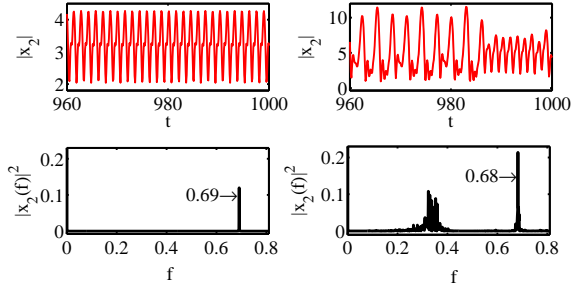}
\caption{(Color online) V system: (Top) The temporal evolution of periodic self-pulsing and chaotic dynamics are shown in left and right panels, respectively. (Bottom) The associated frequency spectral density plots. The operating point $|y_2|=38$ for periodic self-pulsing and $|y_2|=47$ for chaotic behavior, and the other parameters are same as in Fig.~\ref{v_bis_bis_2}.}
\label{v_temp}
\end{figure}  
 
  In general, with variation in the system parameter ($|y_2|$ in the present case) the nonlinear dynamical self-pulsing behavior is expected to occur beyond the Hopf (H) bifurcation point that separates the normal switching from periodic self-pulsing.  Interestingly periodic self-pulsing behavior coexists with stable switching for a range of $|y_2|$ (see region around $|y_2|=37$ shown in Fig.~\ref{v_bif_lyap}). To understand such coexistence we use the limit-cycle continuation from the Hopf point using the MATCONT continuation package~\cite{mat1}. The continuation diagram is shown in Fig.~\ref{v_limit_cont} and indicates that the limit-cycle continuation initially occurs in direction of decreasing $|y_2|$ upto $|y_2|=35$, and then turns around at a Limit-point of cycle (LPC), this is also known {\em i.e} Fold bifurcation point (indicated in cyan color).  
\begin{figure}
\centering
\includegraphics[width=0.5\textwidth]{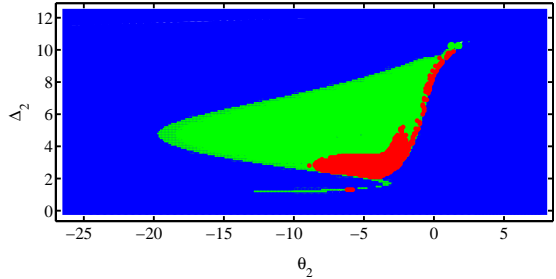}
\caption{(Color online) V system: Stability domain map indicating stable-fixed-point region (blue), the self-pulsing region (green), and the chaotic region (red) as the atomic detuning ($\Delta_2$) and the cavity detuning ($\theta_2$) are varied. The other parameters are as given in Fig.~\ref{v_bis_bis_2} with $|y_1|=15$ and $|y_2|=50$.}
\label{v_domain}
\end{figure}
  The dominant Floquet multipliers plotted on the complex plane indicate that periodic self-pulsing behavior (limit cycles) associated with continuation in the increasing and decreasing variation of $|y_2|$ are stable and unstable, respectively. Therefore, all the three possible behavior (normal switching and stable and unstable periodic self-pulsing) coexist in the lower AOB branch region within the interval $|y_2|$=[35:36]. Hence, depending on the initial condition the appropriate behavior will be observed. The period doubling route described in the above bifurcation diagram is also corroborated by the presence of the period doubling (PD) cycle indicated as cyan loop in Fig.~\ref{v_limit_cont}. 
  \begin{table}
\begin{tabular}{|c|c|c|}
\hline 
\bf Parameters  & \bf Symbol & \bf Range\\
\hline \hline
Atomic detuning &  $\Delta_{1,2}$  & $-10 \leftrightarrow 10$   \\
\hline
Cavity decays &  $\kappa_{1,2}$  & $0 \leftrightarrow 1$  \\
\hline
Cavity detuning &  $\theta_{1,2}$ & $-10 \leftrightarrow 10$  \\
\hline
Cooperative parameters &  $C_{1,2}$ & $100 \leftrightarrow 1000$ \\
\hline
Input field strengths &  $|y_{1,2}|$  & $0 \leftrightarrow 150$  \\
\hline
Non-radiative decay &  $\nu$  & $0 \leftrightarrow 1$  \\
\hline
Incoherent pumping rate &  $r$  & $0 \leftrightarrow 1$  \\
\hline
\end{tabular}
\caption{ All the frequency parameters are scaled with $\gamma_2=36$MHz. }
\label{table2}
\end{table} 
  In the Fig.~\ref{v_temp}, we present the temporal evolution of cavity output field corresponding to stable periodic self-pulsing and chaotic operating regimes, and the corresponding frequency spectrum in the bottom panels. The domain map plotted between the detunings $\theta_2$ and $\Delta_2$ (see Fig.~\ref{v_domain}) portrays the various regions such as normal switching (blue), periodic self-pulsing (green) and chaotic behavior (red) associated with cavity output.

  Before we conclude, we indicate the possible experimental system, where these effects can be realized, using Rubidium ($^{87}$Rb  D$_1$ line) atomic vapor in  the $\Lambda$ and V-type configuration along the transitions $5S_{1/2} (F=2)\leftrightarrow 5P_{1/2}(F'=2) \leftrightarrow 5S_{1/2}(F=1)$ and $5P_{3/2} \leftrightarrow 5S_{1/2} \leftrightarrow 5P_{1/2}$, respectively. Considering  temperatures of about $60^o c$ one would obtain a number density of $\approx 10^{11}$ atoms/$cm^3$ and with transmission coefficient $T \approx 10^{-2}$ would result in cooperative parameter $C \approx 1000$. The input power levels can be varied from $\approx 0 - 20~ mW$ across a spot size of $100 ~\mu m$ to observe effects we have mentioned in this paper. The different values of cooperative parameter can be realized by changing either the number density of atoms or the transmission coefficient of the cavity. The range of parameters are explicitly enumerated in the Table~\ref{table2}. The $4D_{3/2} \leftrightarrow 5P_{3/2} \leftrightarrow 4D_{5/2}$ transitions in $^{87}$Rb offer promising implementation of these effects in the optical communication wavelength range of $1.5\mu m$ with degenerate fields~\cite{Han}.

\section{Conclusion and remarks}
We have demonstrated methods of generating periodic self-pulsing and chaotic instability using double-cavity AOB system having three-level atomic medium in the $\Lambda$ and V configurations. These instabilities occur in the lower branch of the bistable curve and are intrinsically related to the competition of cooperative behavior of the atomic collection along the two adjacent transitions as well as the details of the incoherent pathways within the atom. The ground state decoherence is sufficient for the $\Lambda$ system to exhibit instability, however, by introducing appropriate incoherent pathways (incoherent pump) we induce instability even in the V system double-cavity AOB. Hence, the incoherent pathway needs to deplete the collection of atoms involved in the cooperative effect.

\section{Acknowledgement}
We gratefully acknowledge Dr. Pankaj Wahi for discussions related to nonlinear dynamical aspects of this work.

\end{document}